\documentclass[11pt]{amsart}
\usepackage{amscd}
\usepackage{amsmath}
\usepackage{graphicx}
\usepackage{amsfonts}
\usepackage{amssymb}
\begin{document}

\title[Coherence convertibility for mixed states]
{Coherence convertibility for mixed states}

\author{Xiaofei Qi}
\address[Xiaofei Qi]{Department of Mathematics, Shanxi
University, Taiyuan 030006, P. R.
China}\email{xiaofeiqisxu@aliyun.com}

\author{Zhaofang Bai}
\author{Shuanping Du}
\address[Zhaofang Bai, Shuanping Du]{School of Mathematical
Sciences, Xiamen University, Xiamen 361000,
China}\email{baizhaofang@xmu.edu.cn(corresponding author);
shuanpingdu@yahoo.com}


\begin{abstract}

In this paper, by providing a class of coherence measures in
finite dimensional systems,  a sufficient and necessary condition
for the existence of coherence transformations that convert one
probability distribution of any pure states into another one  is
obtained.
\end{abstract}

\maketitle

\section{Introduction}

Coherence is a fundamental aspect of quantum physics that
encapsulates the defining features of the theory \cite{L}, from the
superposition principle to quantum correlations. It is a key
component in various quantum information and estimation protocols
and is primarily accountable for the advantage offered by quantum
tasks versus classical ones \cite{GLL,NC}.  It has been shown that a
good definition of coherence does not only depend on the state of
the system, but also depend on the a fixed basis for the quantum
system \cite{BCP}. So far, several themes of coherence have been
considered such as witnessing coherence \cite{LLC}, catalytic
coherence \cite{A} and the thermodynamics of quantum coherence
\cite{RFA}.

But, given a quantum state, how much the coherence does it
contain? How to quantify the quantum coherence? There is no
well-accepted efficient method to quantify the coherence in
quantum system until recently.  Baumgratz et al. \cite{BCP}
introduced a rigorous framework for quantification of coherence
and proposed several measures of coherence, which are based on the
well-behaved metrics including the $l_p$-norm, relative entropy,
trace norm and fidelity. 
After then, the quantification of coherence stimulated a lot of
further considerations (see
\cite{SXFL,XLF,MS,RM,SSDB,SBDP,SSDBA,BD1,BD2,BDQ}). Especially, a
general method to derive a series of coherence measures from
concave functions was given in \cite{BDQ} which plays a key role
in this paper.

In quantum information science, the question what tasks may be
accomplished using a given physical resource  is of fundamental
importance in many areas. It is well known that entanglement is a
useful physical resource in many processes of quantum information
processing. In order to perform some tasks, it is key to
manipulate the entanglement under special conditions, namely
allowing only local operations and classical communication (LOCC).
The celebrated Nielsen theorem exposed the necessary and
sufficient conditions for pure bipartite entanglement
transformations \cite{N}. Then Jonathan and Plenio \cite{JP}
extended this result to the case that a pure bipartite state can
be transformed into a probability distribution of pure states.
Later, Li and Shi \cite{LS} gave necessary and sufficient
conditions that a bipartite mixed state
 can be transformed into another mixed state  by LOCC. Other related
results can be found in \cite{GG,Li,Gh} and the references
therein.

For coherence, in \cite{BCP}, the authors proposed a question
similar to entanglement: whether incoherent operations can
introduce an order on the set of quantum states, i.e., whether,
given any two states $\rho$ and $\sigma$, either $\rho$ can be
transformed into $\sigma$ or vice versa under incoherent
operations. In \cite{BDG}, the authors gave an affirmative answer
to this question for pure states by majorization condition.

In this letter, we are aimed to determine when a mixed state
$\rho$ can be transformed to a mixed state $\sigma$ by incoherent
operations. By constructing new classes of coherence measures
using the method in \cite{BDQ}, we partially answer this question
raised in \cite{BCP}.

This paper is organized as follows. In Section 2, we introduce the
concept of coherence measures and the approach how to construct
general coherence measures. Based on these, two new classes of
coherence measures are given, which are key for our main results.
Section 3 is devoted to obtaining a necessary and sufficient
condition that an ensemble can be  transformed into another one by
incoherent quantum operations. We summarize our results in Section
4.

\section{Coherence measures}

\subsection{The construction of coherence measures}

In this paper, we always assume that $H$ is a finite dimensional
Hilbert space with $\dim H=d$. Let ${\mathcal S}(H)$ be the space
of all states on $H$. Fixing a particular basis
$\{|i\rangle\}_{i=1}^d$, recall that a state $\rho\in{\mathcal
S}(H)$ is called incoherent if $\rho$ is diagonal in the fixed
basis, that is, $\rho=\sum_{i=1}^d\lambda_i|i\rangle \langle i|$
with $\lambda_i\geq 0$ and $\sum_{i=1}^d\lambda_i=1$. Denote by
${\mathcal I}$ the set of all incoherent quantum states in $H$.
Quantum operations are specified by a finite set of Kraus
operators $\{K_n\}$ satisfying $\sum_n K_n^\dag K_n=I$, $I$ is the
identity operator on $H$.  Quantum operations are incoherent (ICO)
if they fulfil $K_n\rho K_n^\dag/Tr(K_n\rho K_n^\dag)\in {\mathcal
I}$ for all $\rho\in {\mathcal I}$ and for all $n$. By \cite{BCP},
any proper measure of the coherence $C$ must satisfy the following
conditions:

(C1) $C(\delta)=0$ for all $\delta\in{\mathcal I}$;

(C2a) Monotonicity under all incoherent operations (ICO) $\Phi$:
$C(\rho)\geq C(\Phi(\rho))$,

or (C2b) Monotonicity for average coherence under sub-selection
based on measurements  outcomes: $C(\rho)\geq\sum_np_nC(\rho_n)$
for all $\{K_n\}$ with $\sum_nK_n^\dag K_n=I$ and  $K_n{\mathcal
I}K_n^\dag\subset{\mathcal I}$, where $\rho_n=\frac{K_n\rho
K_n^\dag}{p_n}$ and $p_n={\rm Tr}(K_n\rho K_n^\dag)$;

(C3) Non-increasing under mixing of quantum states:
$\sum_np_nC(\rho_n)\geq C(\sum_np_n\rho_n)$ for any set of states
$\{\rho_n\}$ and any $p_n\geq 0$ with $\sum_np_n=1$.

Note that the conditions (C2b) and (C3) automatically imply the
condition (C2a). The reason we listed all conditions above is that
(similar to entanglement measures) there  exist meaningful
quantifiers of coherence which satisfy the conditions (C1), (C2a)
and (C3), but for which the condition (C2b) is  violated (see
\cite{SXFL}).

Ref. \cite{BCP} gave several coherence measures for finite
dimensional systems, which are based on the well-behaved metrics
such as the $l_1$-norm, relative entropy, trace norm and fidelity.
Recently, Du, Bai and Qi \cite{BDQ} gave a general approach by
convex roof to construct coherence measures  for finite
dimensional systems. The relative entropy coherence measure and
$l_1$ norm coherence measure can be derived from the approach.

Let  $\Omega=\{\mathbf{x}=(x_1, x_2,\cdots,x_d)^t\mid \sum_{i=1}^d
x_i=1 \text{ and } x_i\geq 0 \}$, where $(x_1, x_2,\cdots,x_d)^t$
denotes the transpose of row vector $(x_1, x_2,\cdots,x_d)$. It is
not difficult to check that $\Omega$ is a closed set in ${\mathbb
R}^d$. Assume that
 $f:\Omega\mapsto {\mathbb R}^{+}$ is any nonnegative function
 satisfying the following conditions:

(i) $f((1,0 ,\cdots, 0)^t)=0$;

(ii) $f$ is invariant under any permutation transformation
$P_{\pi}$(here, $\pi$ is a permutation of $\{1,2,\cdots, d\}$ and
$P_{\pi}$ is the permutation matrix corresponding to $\pi$):
$f(P_{\pi}\mathbf{x})=f(\mathbf{x})$ for every
$\mathbf{x}\in\Omega$;

(iii) $f$ is concave: $f(\lambda \mathbf{x}+(1-\lambda)
\mathbf{y})\geq \lambda f(\mathbf{x})+(1-\lambda)f( \mathbf{y})$
for all $\lambda\in [0,1]$ and all $
\mathbf{x},\mathbf{y}\in\Omega$.

For any pure state $|\psi\rangle\langle\psi|$ with $|\psi\rangle
=\sum_{i=1}^d\psi_i|i\rangle\in H$, define
$$C_f(|\psi\rangle)=f((|\psi_1|^2, |\psi_2|^2, \cdots
,|\psi_d|^2)^t).$$For any mixed state $\rho$, define
$$C_f(\rho)=\min _{p_i,|\psi_i\rangle}\{\sum_i
p_iC_f(|\psi_i\rangle):\ \rho=\sum\limits_i
p_i|\psi_i\rangle\langle\psi_i|\}.\eqno(1)$$ Then such $C_f$ is a
coherence measure satisfying (C2b) by \cite[Theorem 1]{BDQ}.

\subsection{New coherence measures}

For any vector $ \mathbf{x}=(x_1,x_2,\cdots,x_d)^t\in \Omega$, let
$\pi$ be the permutation of $\{1,2,\cdots,d\}$ such that
$x_{\pi(1)}\geq x_{\pi(2)}\geq\ldots\geq x_{\pi(d)}$. Define a
function $f_l:\Omega\rightarrow{\mathbb R}^+$ by
$$f_l(\mathbf{x})=\sum_{i=l}^dx_{\pi(i)},\ \
l=2,3,\cdots,d.$$  It is easily checked that $f_l$ ($2\leq l\leq d$)
fulfills the conditions (i) and (iii). Also note that any
permutation does not change elements of the vector. So (ii) is also
satisfied by $f_l$ for $2\leq l\leq d$. Thus, by \cite{BDQ}, for
each $l\in\{2,3,\cdots,d\}$, $C_{f_l}$ is a coherence measure.

More generally, for any vector
$\mathbf{x}=(x_1,x_2,\cdots,x_d)^t\in \Omega$, we can define
another function $g_{l,k}:\Omega\rightarrow{\mathbb R}^+$ by
$$g_{l,k}(\mathbf{x})=\sum_{i=l}^d\frac{x_{\pi(i)}}{k}\wedge 1,\ \
l=2,\cdots,d, \ \ \forall k\in(0,1],$$ here
$\frac{x_{\pi(i)}}{k}\wedge 1$ denotes the minimal value of
$\frac{x_{\pi(i)}}{k}$ and $1$. It is clear that each $g_{l,k}$
fulfills the conditions (i) and (ii). Since the following equation
$$(\lambda a+(1-\lambda)b)\wedge 1\geq \lambda(a\wedge 1)+(1-\lambda)(b\wedge 1)$$
holds for any $a,b\geq 0$ and any $\lambda\in[0,1]$, one can show
that $g_{l,k}$ is concave, that is, $g_{l,k}$ satisfies the
condition (iii) for  $l\in\{2,3,\cdots,d\}$ and for any
$k\in(0,1]$. So, by \cite{BDQ} again, each $C_{g_{l,k}}$ is also a
coherence measure. Particularly, if $k=1$, then
$C_{g_{l,1}}=C_{f_l}$.

\section{Convertibility between mixed states}

\subsection{Convertibility between pure states}

In \cite{BCP}, the authors proposed a question: whether incoherent
operations can introduce an order on the set of quantum states,
i.e., whether, given two states $\rho$ and $\sigma$, either $\rho$
can be transformed into $\sigma$ or vice versa. In \cite{BDG}, the
authors gave an affirmative answer by majorization to the question
for pure states, and they proved that, {\it for any unit vectors
$|\phi\rangle=\sum_{i=1}^d \phi_i|i\rangle,
|\psi\rangle=\sum_{i=1}^d \psi_i|i\rangle\in H$,
$|\phi\rangle\langle \phi|$ can be transformed to
$|\psi\rangle\langle \psi|$ by using an incoherent operation if
and only if $C_{f_l}(|\phi\rangle\langle\phi|)\geq
C_{f_l}(|\psi\rangle\langle\psi|)$ for all $2\leq l\leq d$.}
Here, the condition $C_{f_l}(|\phi\rangle)\geq
C_{f_l}(|\psi\rangle)$ for all $2\leq l\leq d$ is in fact equivalent
to the condition
$(|\phi_1|^2,|\phi_2|^2,\cdots,|\phi_d|^2)^t\prec(|\psi_1|^2,|\psi_2|^2,\cdots,|\psi_d|^2)^t$,
that is, $\sum_{i=1}^m|\phi_i|^2\leq\sum_{i=1}^m|\psi_i|^2$ for
$1\leq m\leq d-1$.

However, in practical application, people often need to deal with
mixed states rather than pure ones. Thus, a natural problem is
raised: whether or not majorization is a suitable tool for
transformation from one mixed state into another one. If it is not
true, what is the condition that a mixed state $\rho$ can be
transformed into another mixed state $\sigma$ by using an
incoherent quantum operation, that is, $\rho\xrightarrow{\rm
ICO}\sigma$.
 This will be the purpose of the
next subsection.

\subsection{Convertibility between mixed states}

In this subsection, we will discuss the coherence convertibility
between any mixed states.

We first consider the coherence transformation between any two
ensembles. Let $D_1=\{p_j, |\phi_j\rangle\}_{j=1}^m$ and $D_2
=\{q_i, |\psi_i\rangle\}_{i=1}^n$  be any two ensembles. Assume
that, for each $j\in\{1,\cdots,m\}$, there exists an ICO $\Phi_j$
which outputs the pure states $|\psi_i\rangle$ with conditional
probability $t_{ji}$ for all the possible outcome states, that is,
$\Phi_j(|\phi_j\rangle\langle\phi_j|)=\sum_{i=1}^nt_{ji}|\psi_i\rangle\langle\psi_i|$
with $\sum_{i=1}^nt_{ji}=1$. Here $$t_{ji}=\text{tr}
(K_l^{(j)}|\phi_j\rangle\langle\phi_j|K_l^{(j) \dag}),\quad
|\psi_i\rangle=\frac 1 {\sqrt{t_{ji}}}K_l^{(j)}|\phi_j\rangle$$
and $K_l^{(j)}$ are Kraus operators of $\Phi_j$. Thus, the
transformation $\Phi\equiv(\Phi_1,\cdots,\Phi_m)$, defined between
the ensembles $D_1$ and $D_2$, outputs the states $|\psi_i\rangle$
with probability $q_i=\sum_{j=1}^mp_jt_{ji}$, that is,
$$\Phi(\sum_{j=1}^mp_j|\phi_j\rangle\langle\phi_j|)=\sum_{i=1}^nq_i|\psi_i\rangle\langle\psi_i|.$$
If such $\Phi$ exists, we say that $D_1$ can be transformed into
$D_2$ by an incoherent operation, that is,
$D_1=\{p_j,|\phi_j\rangle\}\xrightarrow{{\rm
ICO}}\{q_i,|\psi_i\rangle\}=D_2$. Particularly, if $m=1$, then
$|\phi_1\rangle\xrightarrow{{\rm ICO}}\{q_i,|\psi_i\rangle\}$
implies that there exists an ICO $\Phi_1$ which outputs pure
states $|\psi_i\rangle$ with probability $q_i$.

\if false Now, for any mixed state $\rho$, by the definition of
$C_{g_{l,k}}$, there exists an ensemble $\{p_j, |\phi_j\rangle\}$ of
$\rho$ such that
$C_{g_{l,k}}(\rho)=\sum_jp_jC_{g_{l,k}}(|\phi_j\rangle)$ for
$l=2,\cdots,d$ and $k\in(0,1]$. Particularly,
$C_{f_l}(\rho)=C_{g_{l,1}}(\rho)=\sum_jp_jC_{g_{l,1}}(|\phi_j\rangle)=\sum_jp_jC_{f_l}(|\phi_j\rangle)$
for $l=2,\cdots,d$. Such ensemble attained the minimum in Eq.(1) is
called an optimal one of $\rho$ in this paper. Assume that $\rho$
and $\sigma$ are two mixed states. Let $\{p_j, |\phi_j\rangle\}$ and
$\{q_i, |\psi_i\rangle\}$ be two optimal ensembles of $\rho$ and
$\sigma$, respectively.

{\bf Definition 1.} $\rho\xrightarrow{\rm ICO}\sigma \text{ iff }
\{p_j,|\phi_j\rangle\}\xrightarrow{{\rm
ICO}}\{q_i,|\psi_i\rangle\}.$

{\bf Remark 2.} In the case of both $\rho$ and $\sigma$ are pure
states, the above definition is equivalent that there is an ICO
$\Phi$ such that $\Phi(\rho)=\sigma$. But in other cases, things are
not so. For example, assume that $\rho$ is pure, $\sigma$ is mixed
and there is an ICO $\Phi$ such that $\Phi(\rho)=\sigma$. While the
$\Phi$ corresponds an ensemble $\{p_i, |\psi_i\rangle\}$ such that
$\rho \xrightarrow{\rm ICO} \{q_i,|\psi_i\rangle\}$, the key lies in
$\{p_i, |\psi_i\rangle\}$ may not be an optimal ensemble. \fi

The following first result gives a necessary and sufficient
condition of coherence transformations between pure states and any
emsembles.

{\bf Theorem 1.} {\it For any pure state $|\phi\rangle\langle\phi|$
and any ensemble $\{p_j,|\psi_j\rangle\}_{j=1}^m$,
$|\phi\rangle\langle\phi|\xrightarrow{{\rm
ICO}}\{p_j,|\psi_j\rangle\}_{j=1}^m$ if and only if
$C_{f_l}(|\phi\rangle)\geq \sum_{j=1}^mp_jC_{f_l}(|\psi_j\rangle)$
for $l=2,\cdots,d$.}

{\bf Proof.} ``$\Rightarrow$": Assume that there exists some
incoherent quantum operation $\Phi$ such that
$\Phi(|\phi\rangle\langle\phi|)=\sum_{j=1}^mp_j|\psi_j\rangle\langle\psi_j|$.
Since $C_{f_l}$ is a coherence measure by Section 2, (C2b) implies
$C_{f_l}(|\phi\rangle)\geq\sum_{j=1}^mp_jC_{f_l}(|\psi_j\rangle)$
for $l=2,\cdots,d$.

``$\Leftarrow$": Assume that $C_{f_l}(|\phi\rangle)\geq
\sum_{j=1}^mp_jC_{f_l}(|\psi_j\rangle)$ for $l=2,\cdots,d$. Write
$|\phi\rangle=\sum_{i=1}^d\phi_{i}|i\rangle$ and
$|\psi_j\rangle=\sum_{i=1}^d\psi_{ji}|i\rangle$. For the
convenience, without loss of generality, we may require that the
coefficients of $|\phi\rangle$ and $|\psi_i\rangle$ are all in the
descending order. Thus, by the definition of $C_{f_l}$, we have
$$\begin{array}{rl}
\sum_{i=l}^d|\phi_i|^2=&C_{f_l}(|\phi\rangle)\geq\sum_{j=1}^mp_jC_{f_l}(|\psi_j\rangle)\\
=&\sum_{j=1}^mp_j\sum_{i=l}^d|\psi_{ji}|^2\\
=&\sum_{i=l}^d\sum_{j=1}^m|\sqrt{p_j}\psi_{ji}|^2.
\end{array}\eqno(2)$$
Define a vector $|\eta\rangle\in H$ by
$$|\eta\rangle=\sum_{i=1}^d\sqrt{\sum_{j=1}^m|\sqrt{p_j}\psi_{ji}|^2}|i\rangle=\sum_{i=1}^d\eta_i|i\rangle.$$
Note that
$$\sum_{i=1}^d|\eta_i|^2=\sum_{i=1}^d\sum_{j=1}^m|\sqrt{p_j}\psi_{ji}|^2=\sum_{j=1}^mp_j(\sum_{i=1}^d|\psi_{ji}|^2)=\sum_{j=1}^mp_j=1.$$
So $|\eta\rangle\langle\eta|$ is  a pure state. Moreover, Eq.(2)
implies
$$C_{f_l}(|\phi\rangle)\geq C_{f_l}(|\eta\rangle),\ \ 2\leq l\leq d.$$
By \cite{BDG}, there exists an ICO $\Phi_1$ such that
$$\Phi_1(|\phi\rangle\langle\phi|)=|\eta\rangle\langle\eta|.\eqno(3)$$

Next, for any $1\leq j\leq m$, define
$$A_j=\sum_{i=1}^d \frac{\sqrt{p_j}\psi_{ji}}{\eta_i}|i\rangle\langle i|.$$
It is easy to check that $\sum_{j=1}^mA_j^\dag A_j=I_d$ and
$A_j{\mathcal I}A_j^\dag\subset{\mathcal I}$ for each $j$. So the
map $\Phi_2$ defined by
$\Phi(\cdot)=\sum_{j=1}^mA_j(\cdot)A_j^\dag$ is an ICO.  In
addition, a simple calculation yields
$$A_j |\eta\rangle=p_j|\psi_j\rangle.\eqno(4)$$
Define a new map $\Phi$ by $\Phi=\Phi_2\circ\Phi_1$. It is obvious
that a composition of any two ICOs is still an ICO. So $\Phi$ is an
ICO; moreover, by Eqs.(3)-(4),  $\Phi$ realizes the required
transformation. \hfill$\Box$

\if false By using Theorem 1, a necessary and sufficient condition
between pure states and mixed states can be obtained.

{\bf Theorem 4.} {\it For any pure state $|\phi\rangle\langle\phi|$
and any mixed state $\rho$ in $H$,
$|\phi\rangle\langle\phi|\xrightarrow{{\rm ICO}}\rho$ if and only if
$C_{f_l}(|\phi\rangle)\geq C_{f_l}(\rho)$ for $l=2,\cdots,d$.}

{\bf Proof.} If $|\phi\rangle\langle\phi|\xrightarrow{{\rm
ICO}}\rho$, then $|\phi\rangle\langle\phi|\xrightarrow{{\rm
ICO}}\{p_j,|\psi_j\rangle\}$ for some optimal ensemble
$\{p_j,|\psi_j\rangle\}$ of $\rho$. It follows from Theorem 1 that
$C_{f_l}(|\phi\rangle)\geq \sum_jp_jC_{f_l}(|\psi_j\rangle)=
C_{f_l}(\rho)$ for $l=2,\cdots,d$. Here, the equality use the
optimality assumption.

Now, assume that $C_{f_l}(|\phi\rangle)\geq C_{f_l}(\rho)$ for
$l=2,\cdots,d$. Let $\{p_j,|\psi_j\rangle\}_{j=1}^m$ be an optimal
ensemble of $\rho$. Then
$$C_{f_l}(\rho)=\sum_jp_jC_{f_l}(|\psi_j\rangle),\ \ l=2,\cdots,d,$$
and so $C_{f_l}(|\phi\rangle)\geq \sum_jp_jC_{f_l}(|\psi_j\rangle)$
for $l=2,\cdots,d$. Theorem 1 implies
$|\phi\rangle\langle\phi|\xrightarrow{{\rm
ICO}}\{p_j,|\psi_j\rangle\}$. Hence
$|\phi\rangle\langle\phi|\xrightarrow{{\rm ICO}}\rho$, completing
the proof. \hfill$\Box$ \fi

However, if the pure state $|\phi\rangle\langle\phi|$ in Theorem 1
is replaced by an ensemble, then the coherence measure $C_{f_l}$ is
not enough to characterize the coherence transformations between the
ensembles. In this case, more coherence measures are needed.

\if false Now, we can give the main result in this paper, that is,
the necessary and sufficient condition of coherence transformation
between any two ensembles.\fi

{\bf Theorem 2.} {\it Assume that $D_1=\{p_j,
|\phi_j\rangle\}_{j=1}^m$ and $D_2 =\{q_i, |\psi_i\rangle\}_{i=1}^n$
are any two ensembles. Then $D_1\xrightarrow{{\rm ICO}}D_2$ if and
only if $\sum_{j=1}^mp_jC_{g_{l,k}}(|\phi_j\rangle)\geq
\sum_{i=1}^nq_iC_{g_{l,k}}(|\psi_i\rangle)$ for $l=2,\cdots,d$ and
$k\in(0,1]$.}

To prove the theorem, the following lemma is  needed.

{\bf Lemma 3.} (\cite{LS}) {\it Assume that
$\{p_i\}_{i=1}^n\subset[0,1]$ and $a\in[0,1]$.  Then  a set of
$\{a_i^\prime\}_{i=1}^n$ satisfies  $\sum_{i=1}^np_ia_i^\prime\geq
a$ if and only if there exists a set of $\{a_i\}_{i=1}^n$ such that
$a_i\leq a_i^\prime$ for all $1\leq i\leq n$ and
$\sum_{i=1}^np_ia_i=a$.}

{\bf Proof of Theorem 2.} For the ``only if" part, if
$D_1\xrightarrow{{\rm ICO}}D_2$, by the definition, for each $j$,
there exists an ICO $\Phi_j:M_d\rightarrow M_d$ such that
$\Phi_j(|\phi_j\rangle\langle\phi_j|)=\sum_{i=1}^nt_{ji}|\psi_i\rangle\langle\psi_i|$
with $\sum_{i=1}^nt_{ji}=1$ and $q_i=\sum_{j=1}^mp_jt_{ji}$. Since
$C_{g_{l,k}}$ is a coherence measure by Section 2, (C2b) implies
$C_{g_{l,k}}(|\phi_j\rangle)\geq\sum_{i=1}^nt_{ji}C_{g_{l,k}}(|\psi_i\rangle)$,
and so
$$\sum_{j=1}^mp_jC_{g_{l,k}}(|\phi_j\rangle)\geq\sum_{j=1}^mp_j\sum_{i=1}^nt_{ji}C_{g_{l,k}}(|\psi_i\rangle)=
\sum_{i=1}^nq_iC_{g_{l,k}}(|\psi_i\rangle).$$

For the ``if" part, assume that
$$\sum_{j=1}^mp_jC_{g_{l,k}}(|\phi_j\rangle)\geq
\sum_{i=1}^nq_iC_{g_{l,k}}(|\psi_i\rangle)\ \ {\rm holds \ for}\
l=2,\cdots,d\ \ {\rm and}\ \ k\in(0,1].\eqno(5)$$ Here, we only give
the detailed proof for the case $d=3$ and $m=n=2$ in Eq.(5). For
higher dimensional cases and larger $m,n$, the proof is similar.

Write $|\psi_j\rangle=\sum_{i=1}^3\psi_{ji}|i\rangle$ and
$|\phi_j\rangle=\sum_{i=1}^3\phi_{ji}|i\rangle$, $j=1,2$. For the
convenience,  without loss of generality, we can require
$|\psi_{i1}|^2\geq|\psi_{i2}|^2\geq|\psi_{i3}|^2$ and
$|\phi_{i1}|^2\geq|\phi_{i2}|^2\geq|\phi_{i3}|^2$ for $j=1,2$. Then
Eq.(5) implies
$$p_1(\frac{|\psi_{13}|^2}{k}\wedge1)+p_2(\frac{|\psi_{23}|^2}{k}\wedge1)
\geq
q_1(\frac{|\phi_{13}|^2}{k}\wedge1)+q_2(\frac{|\phi_{23}|^2}{k}\wedge1)\eqno(6)$$and
$$\begin{array}{rl}&p_1(\frac{|\psi_{12}|^2+|\psi_{13}|^2}{k}\wedge1)+p_2(\frac{|\psi_{22}|^2+|\psi_{23}|^2}{k}\wedge1)\\
\geq&
q_1(\frac{|\phi_{12}|^2+|\phi_{13}|^2}{k}\wedge1)+q_2(\frac{|\phi_{22}|^2+|\phi_{23}|^2}{k}\wedge1),\end{array}\eqno(7)$$
where $k\in(0,1]$ is arbitrary.

Assume that we have proved that there exists a set of
$\{t_{ij}\}_{i,j=1,2}\subseteq[0,1]$ such that
$$C_{f_l}(|\psi_i\rangle)\geq
t_{i1}C_{f_l}(|\phi_1\rangle)+t_{i2}C_{f_l}(|\phi_2\rangle)\eqno(8)$$
holds for $l=2,3$ and $i=1,2$, where $\{t_{ij}\}_{i,j=1,2}$
satisfy the conditions:
$$\begin{cases}\begin{array}{ll}t_{11}+t_{12}=t_{21}+t_{22}=1,\\
q_1=p_1t_{11}+p_2t_{21}.\end{array}\end{cases}\eqno(9)$$ Then, by
Theorem 1, for each $i$, there exists an ICO $\Phi_i$ such that
$$\Phi_i(|\psi_i\rangle\langle\psi_i|)=t_{i1}|\phi_1\rangle\langle\phi_1|+t_{i2}|\phi_2\rangle\langle\phi_2|.$$
It follows from the definition in the first paragraph in Section 3.2
that $\{p_1,p_2,|\psi_1\rangle,|\psi_2\rangle\}\xrightarrow{\rm
ICO}\{q_1,q_2,|\phi_1\rangle,|\phi_2\rangle\}$, and so the ``if"
part holds.

Thus, to show the sufficiency, we only need to check the existence
of $\{t_{ij}\}$ satisfying Ineq.(8) and  Eq.(9). Note that, by the
definitions of $C_{f_l}$, Ineq.(8) implies
$$\begin{cases}\begin{array}{ll}(a1):&|\psi_{13}|^2\geq t_{11}|\phi_{13}|^2+(1-t_{11})|\phi_{23}|^2,\\
(a2):&|\psi_{12}|^2+|\psi_{13}|^2\geq
t_{11}(|\phi_{12}|^2+|\phi_{13}|^2)\\&+(1-t_{11})(|\phi_{22}|^2+|\phi_{23}|^2),\\
(b1):&|\psi_{23}|^2\geq
t_{21}|\phi_{13}|^2+(1-t_{21})|\phi_{23}|^2,\\
(b2):&|\psi_{22}|^2+|\psi_{23}|^2\geq
t_{21}(|\phi_{12}|^2+|\phi_{13}|^2)\\&+(1-t_{21})(|\phi_{22}|^2+|\phi_{23}|^2),\end{array}\end{cases}\eqno(10)$$
which are respectively equivalent to the following inequalities:
$$\begin{cases}\begin{array}{ll}
(a1^\prime):&t_{11}\leq
\frac{|\psi_{13}|^2-|\phi_{23}|^2}{|\phi_{13}|^2-|\phi_{23}|^2}\triangleq S_1,\\
(a2^\prime):&t_{11}\leq
\frac{|\psi_{12}|^2+|\psi_{13}|^2-|\phi_{22}|^2-|\phi_{23}|^2}{|\phi_{12}|^2+|\phi_{13}|^2-|\phi_{22}|^2-|\phi_{23}|^2}\triangleq T_1,\\
(b1^\prime):&t_{21}\leq
\frac{|\psi_{23}|^2-|\phi_{23}|^2}{|\phi_{13}|^2-|\phi_{23}|^2}\triangleq S_2,\\
(b2^\prime):&t_{21}\leq
\frac{|\psi_{22}|^2+|\psi_{23}|^2-|\phi_{22}|^2-|\phi_{23}|^2}{|\phi_{12}|^2+|\phi_{13}|^2-|\phi_{22}|^2-|\phi_{23}|^2}\triangleq
T_2.\end{array}\end{cases}\eqno(11)$$ So, in the rest of the paper,
our goal is to prove the existence of $\{t_{ij}\}$ satisfying Eq.(9)
and Ineq.(10) (or (11))  by using Ineqs.(6)-(7) and Lemma 3.

We will complete it by considering several cases.

{\bf Case 1.} $C_{f_l}(|\psi_1\rangle)\geq
C_{f_l}(|\psi_2\rangle)$ and $C_{f_l}(|\phi_1\rangle)\geq
C_{f_l}(|\phi_2\rangle)$ for $l=2,3$.

This case forces to $C_{f_l}(|\psi_2\rangle)\geq
C_{f_l}(|\phi_2\rangle)$ for $l=2,3$. Otherwise, there exists some
$l\in\{2,3\}$, without loss of generality, assume $l=3$, such that
$C_{f_3}(|\psi_2\rangle)< C_{f_3}(|\phi_2\rangle)$, that is,
$|\psi_{23}|^2<|\phi_{23}|^2$. Note that, the assumptions
$C_{f_3}(|\psi_1\rangle)\geq C_{f_3}(|\psi_2\rangle)$ and
$C_{f_3}(|\phi_1\rangle)\geq C_{f_3}(|\phi_2\rangle)$ respectively
implies $|\psi_{13}|^2\geq|\psi_{23}|^2$ and
$|\phi_{13}|^2\geq|\phi_{23}|^2$. Then, taking $k=|\phi_{23}|^2$
in Ineq.(6), one gets
$$1=p_1+p_2>p_1(\frac{|\psi_{13}|^2}{|\phi_{23}|^2}\wedge1)+p_2(\frac{|\psi_{23}|^2}{|\phi_{23}|^2}\wedge1)\geq
q_1+q_2=1,$$ a contradiction. So for $l=2,3$, we have
$$\begin{cases}C_{f_l}(|\psi_1\rangle)\geq C_{f_l}(|\psi_2\rangle)\geq
C_{f_l}(|\phi_2\rangle),\\
C_{f_l}(|\phi_1\rangle)\geq C_{f_l}(|\phi_2\rangle). \end{cases}$$

{\bf Subcase 1.1.} $C_{f_l}(|\psi_1\rangle)\geq
 C_{f_l}(|\psi_2\rangle)\geq C_{f_l}(|\phi_1\rangle)\geq
C_{f_l}(|\phi_2\rangle)$, $l=2,3$.

It is easy to check that $$C_{f_l}(|\psi_i\rangle)\geq t
C_{f_l}(|\phi_1\rangle)+(1-t)C_{f_l}(|\phi_2\rangle)$$ for $i=1,2$
and any $t\in[0,1]$. This implies that Ineq.(10) holds for any
$t_{ij}$. So we only need to choose suitable $t_{ij}$ such that
Eq.(9) holds. Obviously, such $t_{ij}$ are existent.

{\bf Subcase 1.2.} $C_{f_l}(|\psi_1\rangle)\geq
C_{f_l}(|\phi_1\rangle)\geq C_{f_l}(|\psi_2\rangle)\geq
C_{f_l}(|\phi_2\rangle)$, $l=2,3$.

Clearly, $S_1,T_1\geq 1$ in this case. So
Ineq.(11)(a1$^\prime$)-(a2$^\prime$) hold for any
$t_{11}\in[0,1]$.

For $|\psi_2\rangle$, by Lemma 3, there exist
$t_{11},t_{21}\in[0,1]$ such that Eq.(9) and Ineq.(11)(b1$^\prime$)
hold if and only if
$$p_1+p_2\frac{|\psi_{23}|^2-|\phi_{23}|^2}{|\phi_{13}|^2-|\phi_{23}|^2}\geq q_1;\eqno(12)$$
and there exist $t_{11},t_{21}\in[0,1]$ such that Eq.(9) and
Ineq.(11)(b2$^\prime$) hold if and only if
$$p_1+p_2\frac{|\psi_{22}|^2+|\psi_{23}|^2-|\phi_{22}|^2-|\phi_{23}|^2}{|\phi_{12}|^2+|\phi_{13}|^2-|\phi_{22}|^2-|\phi_{23}|^2}\geq q_1.\eqno(13)$$
By taking $k=|\phi_{13}|^2$ and $k=|\phi_{12}|^2+|\phi_{13}|^2$ in
Ineqs.(6)-(7), respectively, and by the assumption, one obtains
$$p_1|\phi_{13}|^2+p_2|\psi_{23}|^2\geq q_1|\phi_{13}|^2+q_2|\phi_{23}|^2\eqno(14)$$and
$$\begin{array}{rl}&p_1(|\phi_{12}|^2+|\phi_{13}|^2)+p_2(|\psi_{22}|^2+|\psi_{23}|^2)\\
\geq
&q_1(|\phi_{12}|^2+|\phi_{13}|^2)+q_2(|\phi_{22}|^2)+|\phi_{23}|^2).\end{array}\eqno(15)$$
A simple calculation gets Ineq.(12)$\Leftrightarrow $Ineq.(14)
and Ineq.(13)$\Leftrightarrow$Ineq.(15).

Now, by taking  $t_{21}\leq\min\{S_0,T_0\}$, the above discussion
guarantees that there exist $t_{11}$ and $t_{21}$ such that Eq.(9)
and Ineq.(11) can be satisfied.

{\bf Subcase 1.3.} $C_{f_l}(|\phi_1\rangle)\geq
C_{f_l}(|\psi_1\rangle)\geq C_{f_l}(|\psi_2\rangle)\geq
C_{f_l}(|\phi_2\rangle)$, $l=2,3$.

For $|\psi_1\rangle$ and $|\psi_2\rangle$, by Lemma 3, there exist
$t_{11},t_{21}\in[0,1]$ such that Eq.(9), Ineqs.(11)(a1$^\prime$)
and (b1$^\prime$) hold if and only if
$$p_1\frac{|\psi_{13}|^2-|\phi_{23}|^2}{|\phi_{13}|^2-|\phi_{23}|^2}+p_2\frac{|\psi_{23}|^2-|\phi_{23}|^2}{|\phi_{13}|^2-|\phi_{23}|^2}\geq q_1,$$
that is,
$$p_1|\psi_{13}|^2+p_2|\psi_{23}|^2\geq q_1|\phi_{13}|^2+q_2|\phi_{23}|^2;\eqno(16)$$
and there exist $t_{11},t_{21}\in[0,1]$ such that Eq.(9) and
Ineqs.(11)(a2$^\prime$) and (b2$^\prime$) hold if and only if
$$\begin{array}{rl}&p_1\frac{|\psi_{12}|^2+|\psi_{13}|^2-|\phi_{22}|^2-|\phi_{23}|^2}{|\phi_{12}|^2+|\phi_{13}|^2-|\phi_{22}|^2-|\phi_{23}|^2}
+p_2\frac{|\psi_{22}|^2+|\psi_{23}|^2-|\phi_{22}|^2-|\phi_{23}|^2}{|\phi_{12}|^2+|\phi_{13}|^2-|\phi_{22}|^2-|\phi_{23}|^2}\geq
q_1,\end{array}$$ that is,
$$\begin{array}{rl}&p_1(|\psi_{12}|^2+|\psi_{13}|^2)+p_2(|\psi_{22}|^2+|\psi_{23}|^2)\\
\geq
&q_1(|\phi_{12}|^2+|\phi_{13}|^2)+q_2(|\phi_{22}|^2)+|\phi_{23}|^2).\end{array}\eqno(17)$$
By taking $k=1$ in Ineqs.(6)-(7),  one can get Ineqs.(16) and
(17). So, by taking $t_{11}\leq\min\{S_1,T_1\}$ and
$t_{21}\leq\min\{S_2,T_2\}$, the above discussion guarantees that
there exist $t_{11}$ and $t_{21}$ such that Eq.(9) and Ineq.(11)
can be satisfied.

{\bf Subcase 1.4.} $C_{f_i}(|\psi_1\rangle)\geq
C_{f_i}(|\phi_1\rangle)$ and
$C_{f_j}(|\psi_1\rangle)<C_{f_j}(|\phi_1\rangle)$ for
$i\not=j\in\{2,3\}$.

In this case, we have either
$$\begin{cases}\begin{array}{ll}C_{f_l}(|\psi_1\rangle)\geq C_{f_l}(|\psi_2\rangle)\geq
C_{f_l}(|\phi_2\rangle),\\
C_{f_l}(|\phi_1\rangle)\geq C_{f_l}(|\psi_2\rangle)\geq
C_{f_l}(|\phi_2\rangle);\end{array}\end{cases}\eqno(18)$$ or
$$\begin{cases}\begin{array}{ll}C_{f_l}(|\psi_1\rangle)\geq C_{f_l}(|\psi_2\rangle)\geq
C_{f_l}(|\phi_2\rangle),\\
C_{f_l}(|\phi_1\rangle)\geq
C_{f_l}(|\phi_2\rangle),\\
C_{f_i}(|\psi_2\rangle)\geq C_{f_i}(|\phi_1\rangle),\\
C_{f_j}(|\psi_2\rangle)<C_{f_j}(|\phi_1\rangle).\end{array}\end{cases}\eqno(19)$$

Without loss of generality, assume that $i=2$ and $j=3$. We only
deal with the case (18). For the case (19), the proof is similar
to that of Eq.(18).

Obviously, $T_1\geq 1$, and so Ineq.(11)(a2$^\prime$) holds for
any $t_{11}$. Thus, there exist $t_{11},t_{21}\in[0,1]$ such that
Eq.(9), Ineq.(11)(a1$^\prime$) and (b1$^\prime$) hold if and only
if Ineq.(16) holds;  there exist $t_{11},t_{21}\in[0,1]$ such that
Eq.(9) and Ineq.(11)(b2$^\prime$) hold if and only if Ineq.(13)
holds. By taking $k=|\phi_{13}|^2$ and
$k=|\phi_{12}|^2+|\phi_{13}|^2$ in Ineqs.(6)-(7), respectively,
Ineq.(16) and Ineq.(13) hold. Hence suitable $t_{11}$ and $t_{21}$
satisfying Eq.(9) and Ineq.(11) exist.

{\bf Subcase 1.5.} $C_{f_i}(|\psi_2\rangle)\geq
C_{f_i}(|\phi_1\rangle)$ and
$C_{f_j}(|\psi_2\rangle)<C_{f_j}(|\phi_1\rangle)$ for
$i\not=j\in\{2,3\}$.

The proof is similar to that of Subcase 1.4.

{\bf Case 2.} $C_{f_i}(|\psi_1\rangle)\geq
C_{f_i}(|\psi_2\rangle)$ and
$C_{f_j}(|\psi_1\rangle)<C_{f_l}(|\psi_2\rangle)$ for
$i\not=j\in\{2,3\}$.

{\bf Case 3.} $C_{f_i}(|\phi_1\rangle)\geq
C_{f_i}(|\phi_2\rangle)$ and
$C_{f_j}(|\phi_1\rangle)<C_{f_l}(|\phi_2\rangle)$ for
$i\not=j\in\{2,3\}$.

For the proofs of Case 2 and Case 3 are similar to that of Case 1.
We omit it here.

Combining Cases 1-3, the proof of the theorem is finished.
\hfill$\Box$

{\bf Remark 4.} Note that, for any irrational number $k$, there
always exist two series of rational numbers
$\{\mu_n,\nu_n\}_{n=1}^\infty\subset(0,1]$ with $\mu_n\leq
k\leq\nu_n$ for each $n$ such that
$\lim_{n\rightarrow\infty}\mu_n=\lim_{n\rightarrow\infty}\nu_n=k$.
Thus, if necessary, by  making some slight modifications  in the
proof of Theorem 2, it is enough to require that
$\sum_{j=1}^mp_jC_{g_{l,k}}(|\phi_j\rangle)\geq
\sum_{i=1}^nq_iC_{g_{l,k}}(|\psi_i\rangle)$ ($l=2,\cdots,d$) for all
rational numbers $k\in(0,1]$.

Finally, we discuss the transformation between any mixed states. For
any mixed state $\rho$, we call an ensemble $\{p_j,
|\phi_j\rangle\}$ of $\rho$ optimal if it attains the minimum in
Eq.(1), that is,
$C_{g_{l,k}}(\rho)=\sum_jp_jC_{g_{l,k}}(|\phi_j\rangle)$ for
$l=2,\cdots,d$ and $k\in(0,1]$. If such optimal ensemble for any
mixed states exists, we can give the definition of coherence
convertibility between mixed states by ICO.

{\bf Definition 5.} $\rho\xrightarrow{\rm ICO}\sigma \text{ iff }
\{p_j,|\phi_j\rangle\}\xrightarrow{{\rm
ICO}}\{q_i,|\psi_i\rangle\}.$ Here, $\{p_j, |\phi_j\rangle\}$ and
$\{q_i, |\psi_i\rangle\}$ are two optimal ensembles of $\rho$ and
$\sigma$, respectively.

{\bf Remark 6.} In the case of both $\rho$ and $\sigma$ are pure
states, the above definition is equivalent that there is an ICO
$\Phi$ such that $\Phi(\rho)=\sigma$. But in other cases, things are
not so. For example, assume that $\rho$ is pure, $\sigma$ is mixed
and there is an ICO $\Phi$ such that $\Phi(\rho)=\sigma$. While the
$\Phi$ corresponds an ensemble $\{p_i, |\psi_i\rangle\}$ such that
$\rho \xrightarrow{\rm ICO} \{q_i,|\psi_i\rangle\}$, the key lies in
$\{p_i, |\psi_i\rangle\}$ may not be an optimal ensemble.

However, we do not know whether an optimal ensemble for any mixed
state exists by now. If there is such ensemble, by Theorem 2 and
Definition 5, the following result is immediate.

{\bf Theorem 7.} {\it Assume that $\rho,\sigma\in{\mathcal S}(H)$
are any two mixed states. Then $\rho\xrightarrow{{\rm ICO}}\sigma$
if and only if $C_{g_{l,k}}(\rho)\geq C_{g_{l,k}}(\sigma)$ for
$l=2,\cdots,d$ and $k\in(0,1]$.}

\if false Unfortunately, $C_{{g_{l,k}}}$ is not a necessary and
sufficient condition for coherence convertibility between mixed
states.

{\bf Theorem 5.} {\it For $d\geq 3$, there exist density matrices
$\rho$ and $\sigma$ such that (i) $C_{g_{l,k}}(\rho)\geq
C_{g_{l,k}}(\sigma)$ for $l=2,\cdots,d$ and $k\in(0,1]$, and (ii)
there is no incoherent quantum operation $\Phi$ such that
$\Phi(\rho)=\sigma$, that is, $\rho\not\xrightarrow{\rm
ICO}\sigma$.}

{\bf Proof.} We first assume that $d\geq 4$. Take three numbers
$\lambda_1,\lambda_2,\lambda_3$ with
$\frac{1}{2}>\lambda_1>\lambda_2>\lambda_3>0$. Assume that
$|\phi_1\rangle=\sqrt{\lambda_1}|1\rangle+\sqrt{1-\lambda_1}|2\rangle$,
$|\phi_2\rangle=\sqrt{\lambda_2}|3\rangle+\sqrt{1-\lambda_2}|4\rangle$
and
$|\psi\rangle=\sqrt{\lambda_3}|1\rangle+\sqrt{1-\lambda_3}|2\rangle$.
Define
$\rho=p_1|\phi_1\rangle\langle\phi_1|+p_2|\phi_2\rangle\langle\phi_2|$
with $p_1,p_2\in[0,1]$ and $p_1+p_2=1$, and
$\sigma=|\psi\rangle\langle\psi|$. Then $\rho,\sigma\in{\mathcal
S}(H)$. By the definition of $C_{g_{l,k}}$, we have
$$C_{g_{2,k}}(|\phi_i\rangle)=\frac{\lambda_i}{k}\wedge 1,\ \
C_{g_{l,k}}(|\phi_i\rangle)=0\ \ {\rm for}\ \ l=3,\cdots,d  \ \ {\rm
and}\ \ i=1,2$$ and
$$C_{g_{2,k}}(|\psi\rangle)=\frac{\lambda_3}{k}\wedge 1,\ \ C_{g_{l,k}}(|\psi\rangle)=0\ \ {\rm for}\ \ l=3,\cdots,d.$$
A direct calculation yields
$$p_1C_{g_{l,k}}(|\phi_1\rangle)+p_2C_{g_{l,k}}(|\phi_2\rangle)\geq C_{g_{l,k}}(|\psi\rangle)
\ \ {\rm for}\ \ l=2,\cdots,d\ \ {\rm and\ any}\ \
k\in(0,1].\eqno(20)$$ On the other hand, define a map $\Phi$ by
$\Phi(\cdot)=\sum_{i=1}^3K_i(\cdot)K_i^\dag$ with
$K_1=|1\rangle\langle 1|+|2\rangle\langle 2|$, $K_2=|3\rangle\langle
3|+|4\rangle\langle 4|$ and $K_3=\sum_{i=5}^d|i\rangle\langle i|$.
It is easy to see that $\Phi$ is an ICO, and that $K_i\rho
K_i^\dag=p_i|\phi_i\rangle\langle\phi_i|$ for $i=1,2$ and $K_3\rho
K_3^\dag=0$. Thus, for any coherence measure $C$, by (C2b), one gets
$C(\rho)\geq p_1C(|\phi_1\rangle)+p_2C(|\phi_2\rangle)$; by (C3),
one has $C(\rho)\leq p_1C(|\phi_1\rangle)+p_2C(|\phi_2\rangle)$. So
$$C(\rho)=p_1C(|\phi_1\rangle)+p_2C(|\phi_2\rangle)\ \ {\rm for \
any\ coherence\ measure}\ \ C.\eqno(21)$$ Particularly,
$C_{g_{l,k}}(\rho)=p_1C_{g_{l,k}}(|\phi_1\rangle)+p_2C_{g_{l,k}}(|\phi_2\rangle)$,
which and Ineq.(20) imply
$$C_{g_{l,k}}(\rho)\geq
C_{g_{l,k}}(\sigma) \ \ {\rm for}\ \ l=2,\cdots,d\ \ {\rm and\ any}\
\ k\in(0,1].$$

Next, we will show that there is no incoherent quantum operation
$\Phi$ such that $\Phi(\rho)=\sigma$. To do this, we use the measure
of relative entropy coherence $C_{\rm RE}$ defined in \cite{BCP} by
$$C_{\rm RE}(\rho^\prime):=\min_{\sigma^\prime\in{\mathcal
I}}S(\rho^\prime\|\sigma^\prime)=S(\rho^\prime_{\rm
diag})-S(\rho^\prime)$$ for any
$\rho^\prime=\sum_{i,j}p_{ij}|i\rangle\langle j|$, where
$S(\rho^\prime\|\sigma^\prime)={\rm Tr}(\rho^\prime{\rm
log}_2\rho^\prime-\rho^\prime{\rm log}_2\sigma^\prime)$ is relative
entropy, $S(\cdot)$ is the von Neumann entropy (\cite{NC}) and
$\rho^\prime_{\rm diag}=\sum_{i}p_{ii}|i\rangle\langle i|$. Now, we
calculate $C_{\rm RE}(\rho)$ and $C_{\rm RE}(\sigma)$. Note that the
von Neumann entropy of any pure states is zero. By the definitions
of $\rho$ and $\sigma$, we have
$$C_{\rm RE}(|\phi_1\rangle\langle\phi_1|)
=S(|\phi_1\rangle\langle\phi_1|_{\rm
diag})-S(|\phi_1\rangle\langle\phi_1|)
=\lambda_1\log_2\lambda_1+(1-\lambda_1)\log_2(1-\lambda_1),$$
$$C_{\rm RE}(|\phi_2\rangle\langle\phi_2|)
=\lambda_2\log_2\lambda_2+(1-\lambda_2)\log_2(1-\lambda_2)$$ and
$$C_{\rm RE}(|\psi\rangle\langle\psi|)
=\lambda_3\log_2\lambda_3+(1-\lambda_3)\log_2(1-\lambda_3).$$ Define
a function $f(x)=x\log_2x+(1-x)\log_2(1-x)$ for any
$x\in(0,\frac{1}{2})$. Since $f^\prime(x)<0$ and
$\frac{1}{2}>\lambda_1>\lambda_2>\lambda_3>0$, we have
$f(\lambda_1)<f(\lambda_2)<f(\lambda_3)$, which imply
$$p_1C_{\rm RE}(|\phi_1\rangle\langle\phi_1|)+p_2C_{\rm RE}(|\phi_2\rangle\langle\phi_2|)<C_{\rm RE}(|\psi\rangle\langle\psi|)$$
for any $p_1,p_2\in[0,1]$ with $p_1+p_2=1$. It follows from Eq.(21)
that $C_{\rm RE}(\rho)<C_{\rm RE}(|\psi\rangle\langle\psi|)=C_{\rm
RE}(\sigma)$, which implies there does not exist ICO $\Phi$ such
that $\Phi(\rho)=\sigma$.

For the case $d=3$, we take
$\rho=p_1|\phi_1\rangle\langle\phi_1|+p_2|\phi_2\rangle\langle\phi_2|$,
where $p_1,p_2\in[0,1]$ with $p_1+p_2=1$,
$|\phi_1\rangle=\sqrt{\lambda}|1\rangle+\sqrt{1-\lambda}|2\rangle$
and  $|\phi_2\rangle=|3\rangle$ for $\lambda\in(0,1)$. Define a map
$\Phi$ by $\Phi(\cdot)=\sum_{i=1}^2K_i(\cdot)K_i^\dag$ with
$K_1=|1\rangle\langle 1|+|2\rangle\langle 2|$ and
$K_2=|3\rangle\langle 3|$. It is easily checked that $\Phi$ is an
ICO and $K_i\rho K_i^\dag=p_i|\phi_i\rangle\langle\phi_i|$ for
$i=1,2$. So, for any coherence measure $C$, by (C2b), we have
$C(\rho)\geq p_1C(|\phi_1\rangle)+p_2C(|\phi_2\rangle)$; and by
(C3), one gets $C(\rho)\leq
p_1C(|\phi_1\rangle)+p_2C(|\phi_2\rangle)$. Hence
$$C(\rho)=p_1C(|\phi_1\rangle)+p_2C(|\phi_2\rangle)\ \ {\rm for \
any\ coherence\ measure}\ \ C. \eqno(22)$$

Now, define
$\rho=p_1|\phi_1\rangle\langle\phi_1|+p_2|\phi_2\rangle\langle\phi_2|$
and
$\sigma=p_1|\psi_1\rangle\langle\psi_1|+p_2|\psi_2\rangle\langle\psi_2|$,
where $p_1,p_2\in[0,1]$ with $p_1+p_2=1$,
$|\phi_1\rangle=\sqrt{\lambda_1}|1\rangle+\sqrt{1-\lambda_1}|2\rangle$,
$|\psi_1\rangle=\sqrt{\lambda_2}|1\rangle+\sqrt{1-\lambda_2}|2\rangle$
and $|\phi_2\rangle=|\psi_2\rangle=|3\rangle$ for
$\frac{1}{2}>\lambda_1>\lambda_2>0$. It is easy to see that
$C_{g_{2,k}}(|\phi_1\rangle)=\frac{\lambda_1}{k}\wedge 1$,
$C_{g_{2,k}}(|\psi_1\rangle)=\frac{\lambda_2}{k}\wedge 1$ and
$C_{g_{l,k}}(|\phi_2\rangle)=C_{g_{l,k}}(|\psi_2\rangle)=0$ for
$l=2,3$. Then
$$p_1C_{g_{l,k}}(|\phi_1\rangle)+p_2C_{g_{l,k}}(|\phi_2\rangle)
\geq p_1C_{g_{l,k}}(|\psi\rangle)+p_2C_{g_{l,k}}(|\psi_2\rangle)\ \
{\rm for}\ \ l=2,3\ \ {\rm and\ any}\ \ k\in(0,1]. $$ This and
Eq.(22) force to
$$C_{g_{l,k}}(\rho)\geq C_{g_{l,k}}(\sigma)\ \
{\rm for}\ \ l=2,3\ \ {\rm and\ any}\ \ k\in(0,1]. $$

However, by a similar argument to that of the case $d\geq 4$,  one
can show that $C_{\rm RE}(\rho)<C_{\rm RE}(\sigma)$, which also
implies there is no ICO $\Phi$ such that $\Phi(\rho)=\sigma$ in the
case $d=3$. \hfill$\Box$\fi

\section{Conclusion}

In summary, we find a necessary and sufficient condition for the
existence of  transformations that converts an ensemble into
another ensemble. Different from pure states, for determining such
transformation, infinite countable number of conditions based on
coherence measures are required.  We also point out that, if there
exists an optimal ensemble for each mixed state, then the
necessary and sufficient conditions for the existence of coherence
convertibility between any two mixed states can be obtained.  So
we partially answer the question raised by Baumgratz et al.. We
believe that our results will be fruitful in further developments
on convertibility of mixed coherent states.

\vspace{4mm}

{\bf Acknowledgement.}  This work was completed while the authors
were visiting the IQC of the University of Waterloo and Department
of Mathematics and Statistics of the University of Guelph  during
the academic year 2014-2015 under the support of China Scholarship
Council. We  thank Professor David W. Kribs and Professor Bei Zeng
for their hospitality. The research of Qi was partially supported by
the Natural Science Foundation of China (11071249, 11201329) and the
Program for the Outstanding Innovative Teams of Higher Learning
Institutions of Shanxi. The research of Bai and Du was partially
supported by the Natural Science Foundation of China (11001230) and
the Natural Science Foundation of Fujian (2013J01022, 2014J01024).

\end{document}